\documentclass[a4paper,twoside,10pt]{article}
\usepackage{graphicx,publaob}
\usepackage{bm}
\usepackage{amssymb}
\usepackage{amsmath}

\def\papertype{ }
\newcommand{\grad}{{\bf \nabla}}
\newcommand{\PhiN}{\phi}

\newcommand{\Msun}{\ensuremath{\text{M}_\odot}}

\begin{document}

\title{SIMULATING DISK GALAXIES AND INTERACTIONS IN MILGROMIAN DYNAMICS}

\authors{I. THIES$^1$, P. KROUPA$^1$, \lowercase{and} B. FAMAEY$^2$}

\address{$^1$Helmholtz-Institut f\"ur Strahlen- und Kernphysik, D-53115 Bonn}
\Email{ithies}{astro.uni-bonn}{de}
\Email{pavel}{astro.uni-bonn}{de}
\address{$^2$Observatoire astronomique de Strasbourg, F-67000 Strasbourg, France}
\Email{benoit.famaey}{astro.unistra}{fr}
\markboth{SIMULATING DISK GALAXIES IN MILGROMIAN DYNAMICS}{THIES, KROUPA, \& FAMAEY}

\abstract{Since its publication 1983, Milgromian dynamics (aka MOND)
has been very successful in modeling the gravitational potential of galaxies from
baryonic matter alone. However, the dynamical modeling has long been an unsolved
issue. In particular, the setup of a stable galaxy for Milgromian N-body calculations
has been a major challenge. Here, we show a way to set up disc galaxies in
MOND for calculations in the PHANTOM OF RAMSES (PoR) code by L\"ughausen (2015) and
Teyssier (2002). The method is done by solving the QUMOND Poisson equations based on
a baryonic and a phantom dark matter component. The resulting galaxy models are stable
after a brief settling period for a large mass and size range.
Simulations of single galaxies as well as colliding galaxies are shown.}

\section{INTRODUCTION}

Since its invention by Milgrom (1983) modified Newtonian dynamics (MOND) has been
successfully implemented in numerical codes for simulations of galaxies and
galaxy systems (see Famaey \& McGaugh 2012 for a review). The most recent implementation of MOND is the PHANTOM OF RAMSES
(PoR) code by L\"ughausen et al. (2015), based on RAMSES by Teyssier (2002).
In this contribution the setup of stable disk galaxies is described and applied
to the simulation of interacting galaxies. For the first time, the rotation curve
of a galaxy in MOND is calculated.

\section{NUMERICAL METHODS}

The setup of an N-body system in MOND is not as straight-forward as in
Newtonian dynamics since the Milgromian gravitation depends on the Newtonian
acceleration and thus two iterating Poisson equations need to be solved.
The first step is the setup of the desired density profile,
$\rho_\text{b}(\bm x)$. In this contribution exponential disk profiles with a
finite truncation radius are used.
Next the Newtonian Poisson equation is solved to obtain the Newtonian
accelerations, ${\mathbf\grad} \PhiN$. Using the quasi-linear formulation of
MOND (QUMOND, Milgrom 2010), in which the Milgromian
modification term of gravity is represented by a ``phantom dark matter''
density distribution, the combined Newtonian and Milgromian Poisson equation
can be written as
\begin{equation}
\nabla^2 \Phi (\bm x)
= 4 \pi G \rho_\text{b}(\bm x) + 
\nabla \cdot \left[ \nu\left(|{\mathbf\grad} \PhiN|/a_0\right) {\mathbf\grad} \PhiN(\bm x) \right]
\label{eq:poisson}
\end{equation}
or
\begin{equation}
\nabla^2 \Phi (\bm x)
= 4 \pi G \left( \rho_\text{b}(\boldsymbol x) + \rho_\text{ph}(\boldsymbol x) \right) \,.
\label{eq:poisson2}
\end{equation}
Here, $ \rho_\text{b}(\boldsymbol x)$ is the density distribution of the baryonic (real)
matter, and $ \rho_\text{ph}(\boldsymbol x)$ describes the distribution of the
phantom dark matter. By solving the QUMOND Poisson equation the accelerations
and thus the circular velocities of particles in a stable disk galaxy can be
obtained.

The setup is done via the MKGALAXY script by McMillan \& Dehnen (2007), modified
by L\"ughausen in 2015 for MOND. Besides an installation of PoR the script requires
the NEMO stellar dynamics library and the PNBODY Python library for N-body
calculations. Gas initial conditions require to patch the PoR code by
editing the CONDINIT subroutine accordingly. This has been done by I. Thies
in 2015 based on the MERGER patch by D. Chapon in 2010 which is readily
available in the RAMSES installation. The new patch is currently available on
request from the author.


\section{Results}

\begin{figure}
\includegraphics[width=\textwidth]{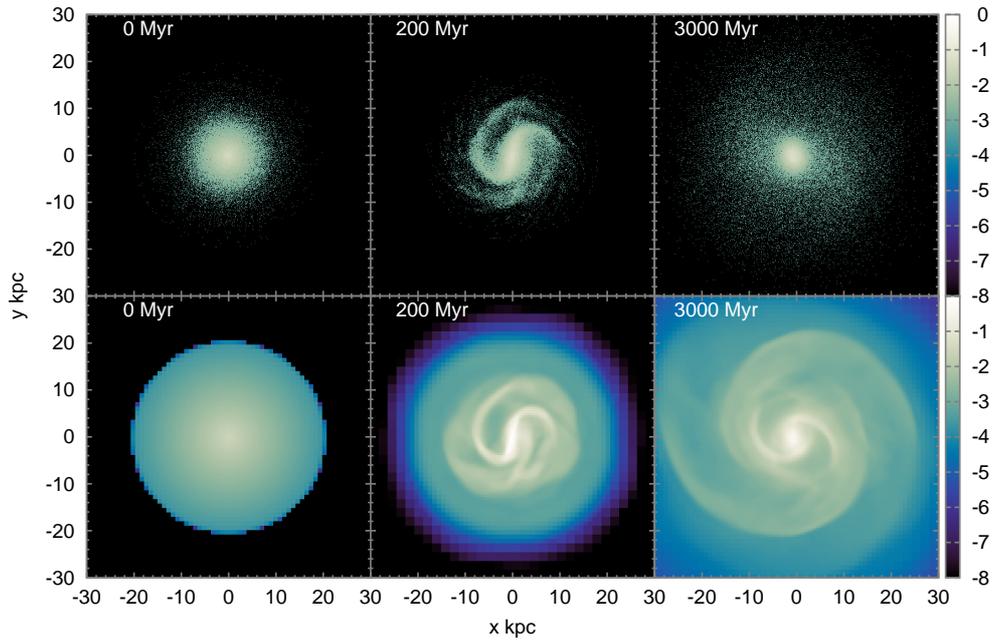}
\caption{\label{isolated}Snapshots of a disk galaxy model at 0, 200 and 3000~Myr after
setup. The upper row shows the stellar component while the lower row
depicts the gas projected density. The colour coding refers to
$\log (\Sigma / [10^9\Msun\,\text{kpc}^{-2}])$. Strong spiral arms form within the first
few 100~Myr and thermalise after a few Gyr with only weak spiral features
remaining in the stellar component. The gas component has spread significantly
but still shows prominent spiral features after 3~Gyr,
when the galaxy has settled completely and does not change much more with time.}
\end{figure}

\begin{figure}
\begin{center}
\includegraphics[width=0.7\textwidth]{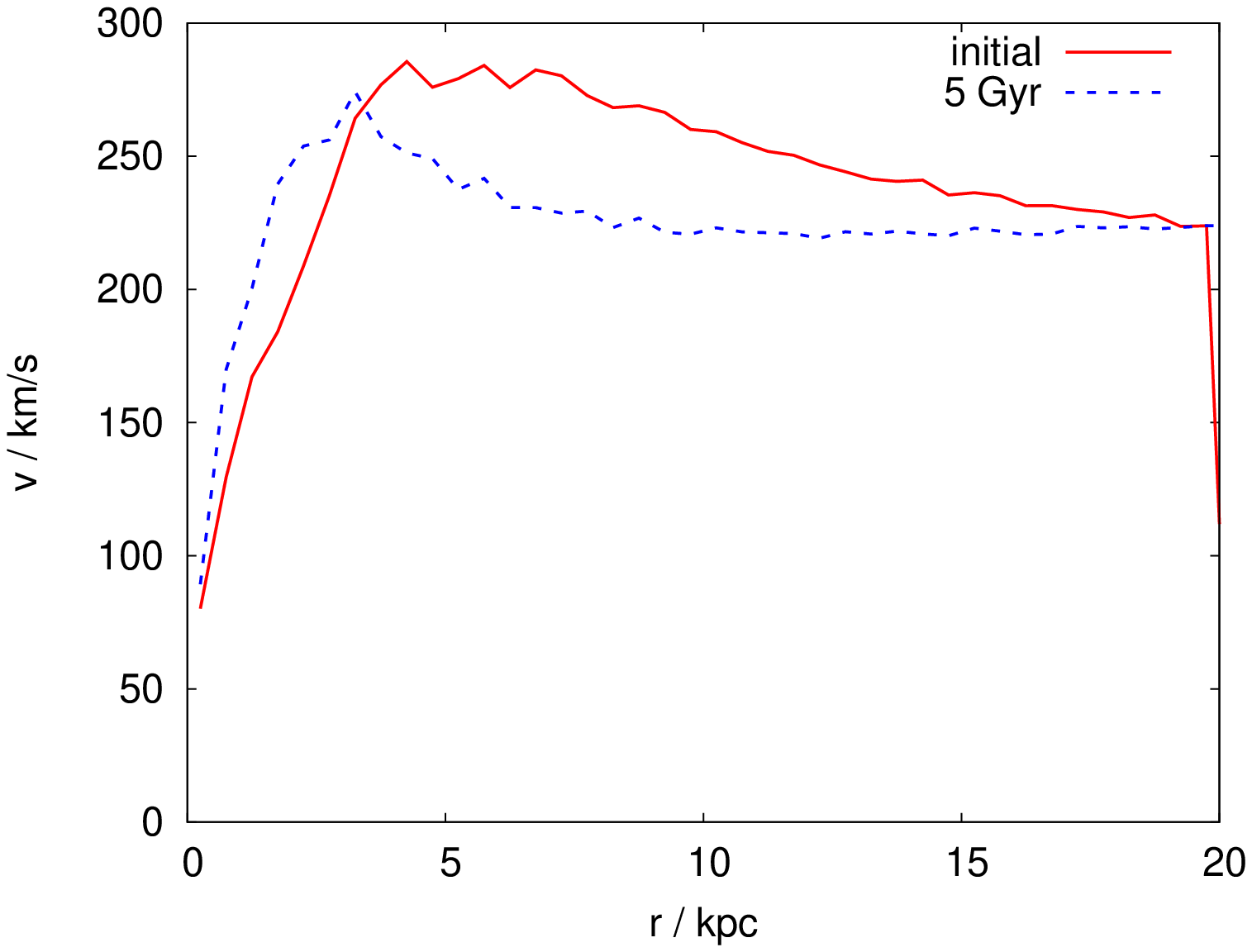}
\caption{\label{rotcurve}The rotation curve at the beginning and
after 5~Gyr. In agreement with observations the rotation curves are relatively
flat in the outer parts.}
\end{center}
\end{figure}

\subsection{Stable isolated galaxies}

A disk galaxy with a stellar mass of $80\cdot10^9\,\Msun$ and gas mass of 
$10\cdot10^9\,\Msun$ is set with an exponential radial profile and a
scale radius of 2~kpc. There is no initial bulge.

In the initial configuration as well as after 5~Gyr the rotation curves are
calculated as the circular velocities from the radial accelerations of the particles.
In agreement with observations the rotation curves are relatively
flat in the outer parts, as can be seen in Fig. \ref{rotcurve}.

\subsection{Interacting galaxies}

\begin{figure}
\begin{center}
\includegraphics[width=\textwidth]{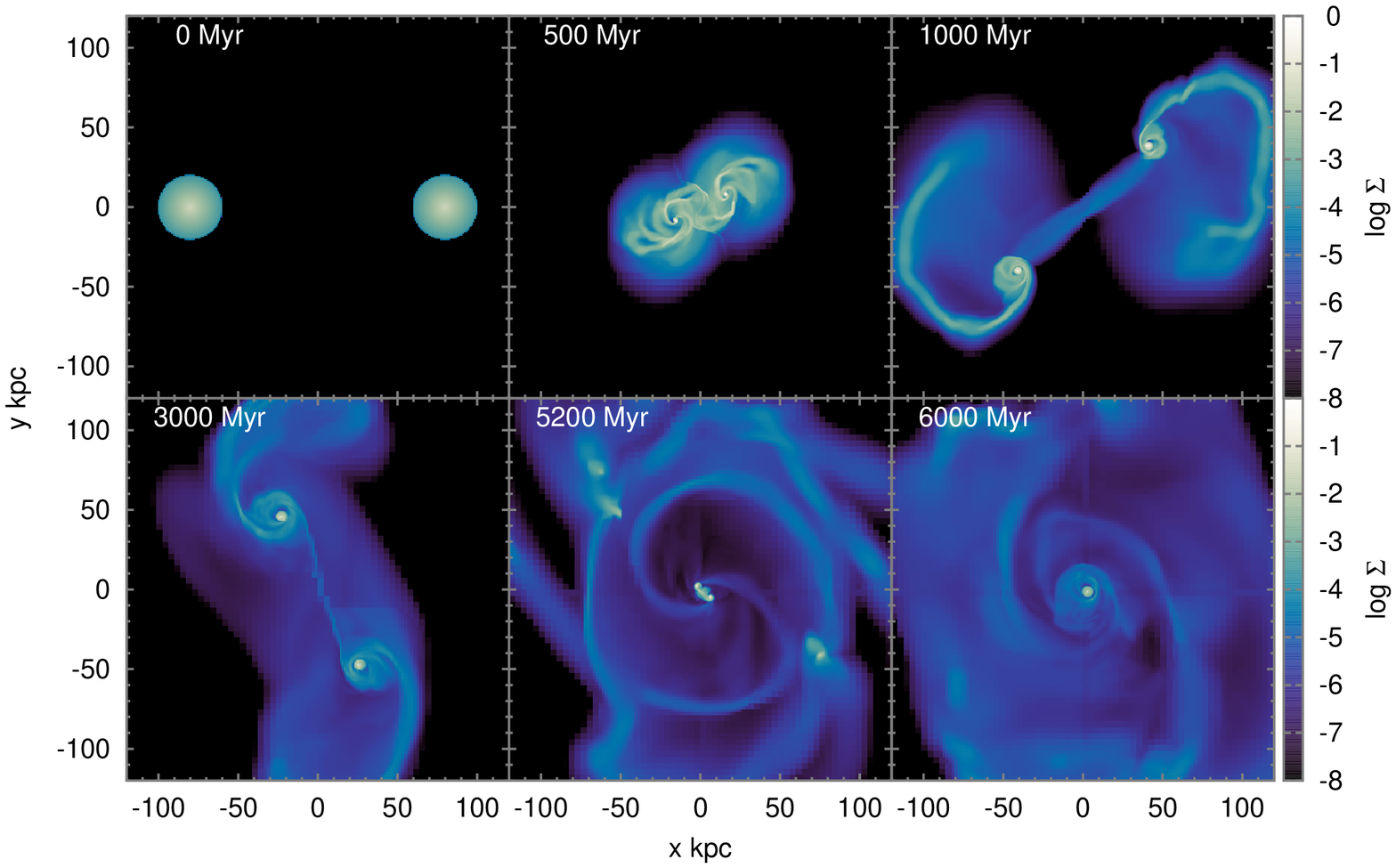}
\caption{\label{merger_gas}Snapshots of the gas component of two interacting galaxies. Note
that it takes about 5500~Myr, measured from the time of collision at timestamp
500~Myr, for the galaxies to merge. This corresponds to six perigalactic passages.
Note the three substantial tidal dwarf galaxies in the lower central panel
(5200~Myr)}
\end{center}
\end{figure}

A pair of galaxies, each with the parameters described above, has been set
for a grazing collision. As can be seen in Fig. \ref{merger_gas}, the interaction
causes prominent tidal arms immediately after collision. The galaxy cores then
orbit each other several times before eventually merging after about 5.5~Gyr.
After about 5~Gyr satellite galaxies are visible as gaseous clumps orbiting
in the plane of the encounter orbit.

\begin{figure}
\begin{center}
\includegraphics[width=0.7\textwidth]{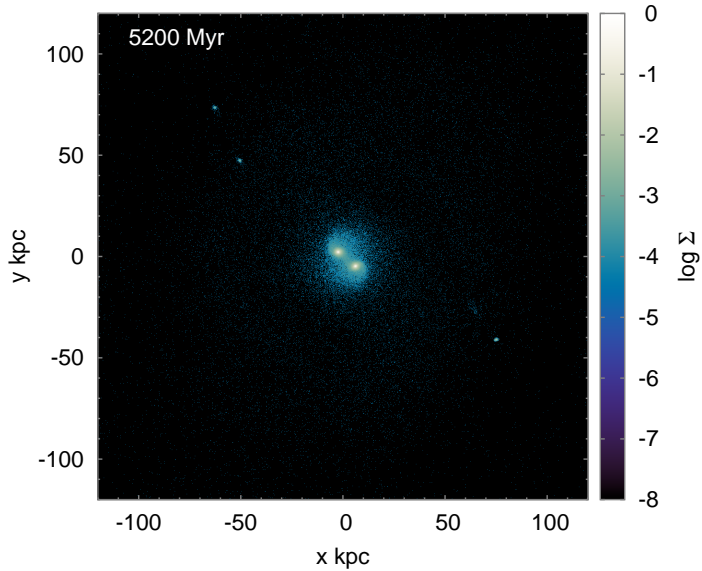}
\caption{\label{merger_stars}Snapshot of the stellar component of two interacting galaxies
at timestamp 5200~Myr. The satellites correspond to the gas clumps in the lower middle
panel in Fig. \ref{merger_gas}.}
\end{center}
\end{figure}

In Fig. \ref{merger_stars} the stellar component of the interacting galaxies
is plotted. More precisely, the stars which formed since the begin of the
simulation are shown in order to emphasise the regions of star formation.
As in the gas plot at timestamp 5200~Myr a few satellite galaxies are visible
around the almost merged original disk galaxies. It has to be noted that
due to the resolution limits of both the gas mesh and the star-representing
particles only the most massive satellites can form and remain stable. In
higher resolution simulations more satellites of lower masses are expected to form.

\section{Summary and future perspectives}

In this contribution the method of setting up stable disk galaxies
in Milgromian dynamics (MOND) has been introduced. For the first time,
the rotation curve of such a galaxy model, which qualitativels matches
observed rotation curves of existing disk galaxies, has been shown.
Furthermore, it has been demonstrated that interacting galaxies tend
to merge relatively late (here after six perigalactic passages within
about 6~Gyr) in contradiction to the $\Lambda$CDM model which predicts
quick mergers due to the dynamical friction of the dark matter halos
(Privon, Barnes et al. 2013).
In addition, the formation of tidal dwarf satellite galaxies in MOND
has been demonstrated.

The ongoing project aims to perform computations with highly increased resolution
in order to reproduce satellite systems like those of the Milky Way Galaxy
and the Andromeda galaxy. Furthermode, the rotation curves of such satellites
are to be calculated. The predictions of these models may then be tested with
observations.

\section{Acknowledgement}

I. Thies and P. Kroupa wish to thank the BELISSIMA team for the invitation.

\references
Famaey, B., McGaugh, S. : 2012, \journal{Liv. Rev. Rel.} 15, 10.

L\"ughausen, F., Famaey, B., Kroupa, P. : 2015, \journal{Can. J. Phys.} \vol{93}, 232.
 
McMillan, P.J., Dehnen, W. : 2007, \journal{MNRAS} \vol{378}, 541.

Milgrom, M. : 1983, \journal{Astrophys. J.} \vol{270}, 365.

Milgrom, M. : 2010, \journal{MNRAS} \vol{403}, 886

Privon, G.C., Barnes, J.E., Evans, A.S. et al. : 2013, \journal{Astrophys. J.} \vol{771}, 120.

Teyssier, R. : 2002, \journal{Astron. Astrophys.} \vol{385}, 337. 
\endreferences

\end{document}